\documentclass[aps,prb,twocolumn,longbibliography]{revtex4-2}
\usepackage{epsfig,amsopn}
\usepackage{graphicx}
\usepackage{color}
\usepackage{amsmath,amssymb}
\usepackage{enumerate}
\newcommand\bea{\begin{eqnarray}}
\newcommand\eea{\end{eqnarray}}
\newcommand\beq{\begin{equation}}
\newcommand\eeq{\end{equation}}

\def\nn{\nonumber}
\def\f{\frac}

\def\de{\delta}

\def\Do{\partial}

\def\th{\theta}

\begin{document}
\title{Transverse response from anisotropic Fermi surfaces} 
\author{ Abhiram Soori}  
\email{abhirams@uohyd.ac.in}
\affiliation{ School of Physics, University of Hyderabad, Prof. C. R. Rao Road, 
Gachibowli, Hyderabad-500046, India}
\begin{abstract}
We demonstrate that an anisotropic and rotated Fermi surface can generate a finite   transverse response in electron transport, even in the absence of a magnetic field or Berry curvature. Using a two-dimensional continuum model, we show that broken $k_y \to -k_y$ symmetry inherent to anistropic bandstructures leads to a nonzero transverse conductivity. We construct a lattice model with direction-dependent nearest- and next-nearest-neighbor hoppings that faithfully reproduces the continuum dispersion and allows controlled rotation of the Fermi contour. Employing a multiterminal geometry and the Büttiker-probe method, we compute the resulting transverse voltage and establish its direct correspondence with the continuum transverse response. The effect increases with the degree of anisotropy and vanishes at rotation angles where mirror symmetry is restored. Unlike the quantum Hall effect, the transverse  response predicted here is not quantized but varies continuously with the band-structure parameters.  Our results provide a symmetry-based route to engineer transverse signals in low-symmetry materials without magnetic fields or topological effects.
\end{abstract}

\maketitle

\section{Introduction}
Free-electron theory provides the simplest description of the electronic properties of metals~\cite{kittel,ashcroft}. While this theory treats electrons as moving in a translationally invariant continuum, the actual electronic motion takes place on a lattice, a feature naturally captured by the tight-binding model. These two descriptions can be mapped onto each other.  Within the free-electron framework, the dispersion takes the familiar parabolic form 
$ E = {\hbar^2 k^2}/{2m} - \mu, $ 
with an effective mass $m$. However, in several classes of materials-including layered semiconductors such as ReSe$_2$~\cite{lin2015,liu2020} the dispersion becomes anisotropic, resulting in direction-dependent effective masses. Even in a two-dimensional electron gas (2DEG) with an originally circular Fermi surface, a series of periodically spaced barrier  potentials can induce anisotropy and distort the Fermi contour~\cite{ismail88,beenakker91}.

The Hall effect, in contrast, appears in several forms. The most familiar is the conventional Hall effect, where a transverse voltage develops in a 2DEG subjected to a perpendicular magnetic field~\cite{kittel,ashcroft}. In the presence of Rashba spin--orbit coupling, an in-plane magnetic field may also generate a transverse response-the planar Hall effect-arising from the field-induced shift of the spin--orbit-coupled Fermi surface~\cite{wadehra20,soori2021}. These examples illustrate that transverse currents and transverse responses need not always originate from Lorentz forces; they can also be rooted in the geometry and symmetry of the Fermi surface.

\begin{figure}
\includegraphics[width=8cm]{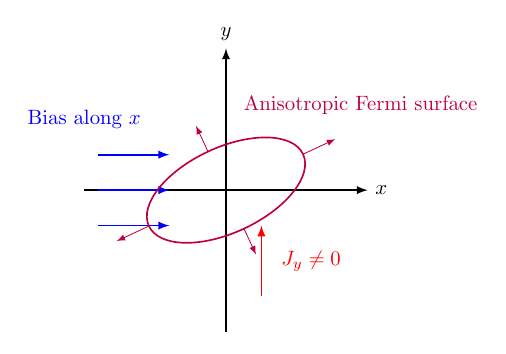}
\caption{ A two-dimensional electron gas with an anisotropic bandstructure, translationally invariant along both directions, is subjected to a bias applied along $\hat{x}$. The elliptical Fermi contour is rotated such that its major axis is misaligned with both $\hat{x}$ and $\hat{y}$. The red arrows denote the direction of the quasiparticle velocity on the contour. As a consequence of this misalignment, a longitudinal bias along $\hat{x}$ generates a net transverse current along $\hat{y}$.}~\label{fig:schem-cont}
\end{figure}

In general, for an isotropic 2DEG with a circular Fermi surface, transverse currents cancel because states with opposite transverse momenta carry equal and opposite contributions. However, this cancellation breaks down when the crystallographic axes of a 2DEG with an anisotropic Fermi surface are rotated relative to the transport direction [see Fig.~\ref{fig:schem-cont}]. In such a configuration, states with opposite transverse momenta and identical longitudinal momentum need not exist, and a net transverse current can arise even in the absence of magnetic fields or broken time-reversal symmetry. Effects rooted in the anisotropy of the Fermi contour have previously been discussed in the context of Josephson junctions~\cite{Sahoo2025socjj,mazumdar2025}.

In this work, we analyse this effect in detail. We first compute the transverse conductivity of a translationally invariant system with an anisotropic Fermi surface. We then examine the emergence of a transverse voltage in a lattice model of a 2DEG using the B\"uttiker-probe technique~\cite{soori2024deco}. While the continuum model yields a transverse conductivity in a translationally invariant setting, the lattice model probes the corresponding transverse voltage in a multiterminal geometry; the latter provides an experimentally relevant manifestation of the same underlying anisotropic transport mechanism. It is known that transverse voltages can arise even in the absence of time-reversal symmetry breaking due to reflection~\cite{song98} or refraction~\cite{ronika2025} induced by spatial inhomogeneities in the potential landscape. Our results demonstrate that anisotropy and crystalline orientation provide a distinct route to generating transverse responses, without relying on magnetic fields, Berry curvature, or spatially varying potentials.

It is useful to contrast the present mechanism with related effects discussed in the literature. In systems with Rashba spin-orbit coupling subject to an in-plane  Zeeman field, the Fermi contour is displaced away from $(k_x,k_y)=(0,0)$, which leads to an imbalance between states with opposite transverse momenta; this underlies the planar Hall effect and related transverse responses in Josephson systems~\cite{soori2021,Sahoo2025socjj}. In contrast, in the work of Song \emph{et al.}~\cite{song98}, the transverse response arises from asymmetric reflection of electrons by engineered antidot structures, while in the work of Sarkar \emph{et al.}~\cite{ronika2025}, it originates from refraction-like bending of electron trajectories due to spatial variations in the potential. In all these cases, the transverse response is tied either to a shift of the Fermi surface or to real-space inhomogeneities. In contrast, the mechanism discussed here relies solely on a rotated anisotropic Fermi surface that remains centred at $(k_x,k_y)=(0,0)$, and therefore does not require any symmetry breaking in real space or momentum-space displacement of the Fermi contour.

\section{Continuum model}

In this section, we describe the system using a continuum model. The system is assumed to be translationally invariant along both $\hat x$ and $\hat y$ directions. Electrons are described within a free--electron framework but with direction-dependent effective masses, leading to an elliptical Fermi contour. The Hamiltonian is

\bea 
H_c &=& -(t-\de\cos{2\phi})a^2\frac{\Do^2}{\Do x^2} - (t+\de\cos{2\phi})a^2\frac{\Do^2}{\Do y^2} \nn \\ && +2\de\sin{2\phi} ~a^2\frac{\Do^2}{\Do x\Do y}, \label{eq:ham-cont}
\eea
where $\de$ denotes the anisotropy parameter and $\phi$ specifies the angle between the major axis of the elliptical Fermi contour and the $\hat x$ direction, and $a$ is a length-scale that can be thought of as lattice spacing of the underlying lattice. Here, $\Do/\partial x$ denotes the spatial derivative, corresponding to the momentum operator $-i\hbar \partial_x$ in the continuum representation. For simplicity, we choose $0 \le \de < t$. In the limiting case $\de=0$, the Fermi contour becomes isotropic, corresponding to an effective mass $\hbar^2/2ta^2$.

The charge current density $\vec J = (J_x,J_y)$ carried by a state with wavevector $\vec k=(k_x,k_y)$ is given by
\bea
J_x &=& 2e[(t-\de\cos{2\phi})k_x - \de\sin{2\phi}~k_y]/\hbar, \nn \\ 
J_y &=& 2e[(t+\de\cos{2\phi})k_y-\de \sin{2\phi}~k_x]/\hbar.
\eea
For a given energy $E$, the wavevector components may be parametrised by an angular parameter $\th$ as
\bea 
k_xa &=& \sqrt{E}\left[ \frac{\cos\th\cos\phi}{\sqrt{t-\de}}
       - \frac{\sin\th\sin\phi}{\sqrt{t+\de}} \right], \nn\\
k_ya &=& \sqrt{E}\left[ \frac{\cos\th\sin\phi}{\sqrt{t-\de}}
       + \frac{\sin\th\cos\phi}{\sqrt{t+\de}} \right], \nn
\eea
with $\th\in[0,2\pi)$. When $\de=0$, the parameter $\th$ reduces to the usual polar angle of $\vec k$ measured from $\hat x$.
It can be shown that $J_x$ is positive for $-\pi/2-\eta < \th < \pi/2-\eta$, where 
\[
\eta = \tan^{-1}\!\left[\tan\phi\sqrt{\frac{t+\de}{t-\de}}\right].
\]
Within the Landauer framework, applying a forward bias populates all states with positive longitudinal velocity. The longitudinal differential conductivity at bias $V=E/e$ is given by

\bea
G_{xx} &=& \frac{e}{8\pi^2a\sqrt{t^2-\de^2}} 
           \int_{-\pi/2-\eta}^{\,\pi/2-\eta} J_x(E,\th)\, d\th. \label{eq:Gxx}
\eea

The transverse differential conductivity quantifies the net transverse current $I_y$ generated by a longitudinal bias $V$. It is defined as $dI_y/dV$, and is given by

\bea
G_{yx} &=& \frac{e}{8\pi^2a\sqrt{t^2-\de^2}}
           \int_{-\pi/2-\eta}^{\,\pi/2-\eta} J_y(E,\th)\, d\th .
\eea

Evaluating the integral yields
\bea
G_{yx} &=& -\frac{e^2}{\pi ha}
\sqrt{\frac{2E(t+\de\cos{2\phi})}{t^2-\de^2}} 
\sin(\eta-\eta'), \label{eq:Gyxc}
\\
&& {\rm where}\quad 
\eta'=\tan^{-1}\!\left[\tan\phi\sqrt{\frac{t-\de}{t+\de}}\right]. \nn
\eea

 \begin{figure}
 \includegraphics[width=8cm]{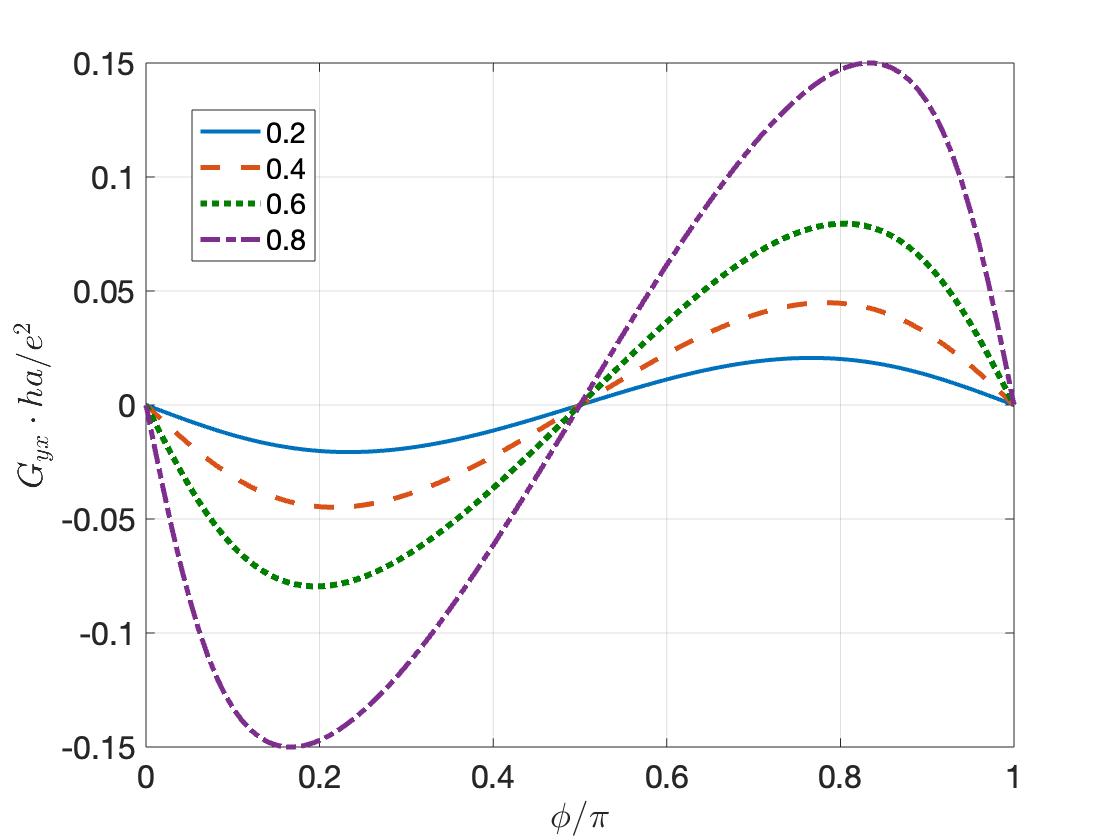}
 \caption{Transverse conductivity as a function of the angle between the major axis of the elliptical Fermi contour and the longitudinal bias direction ($\hat x$) for $E=0.1t$. Different curves correspond to different values of the anisotropy ratio $\delta/t$ (shown in the legend).}\label{fig:Gyx}
 \end{figure}

At $\phi = 0$ and $\phi = \pi/2$, one finds $\eta = \eta'$, and consequently the transverse conductivity vanishes. The behaviour of $G_{yx}$ as a function of $\phi$, the angle between the major axis of the elliptical Fermi contour and the bias direction $\hat x$, is shown in Fig.~\ref{fig:Gyx}.

\section{Lattice Model}

\begin{figure*}
\includegraphics[width=14cm]{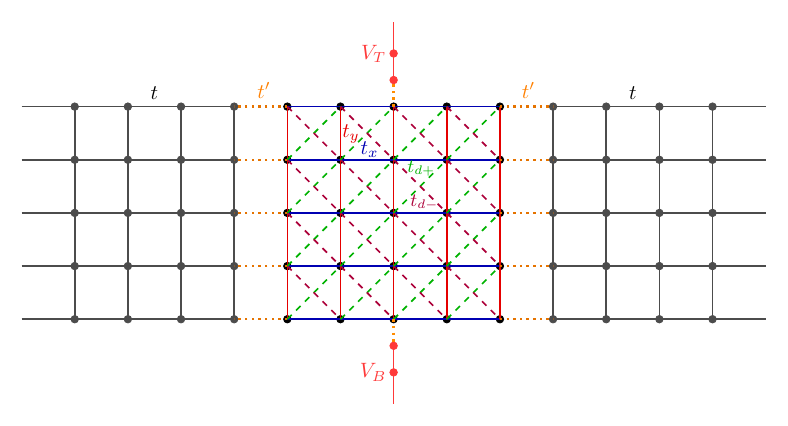}
\caption{
 Schematic of the lattice-based setup used to probe the transverse response arising from an anisotropic Fermi contour. 
The central region is a square lattice with direction-dependent nearest- and next-nearest-neighbor hopping amplitudes, 
connected to source and drain terminals on either side. Two voltage probe terminals are attached symmetrically along 
the transverse direction. A bias voltage $V$ applied between the source and drain drives a longitudinal current, 
while the transverse voltage difference $V_H$ that develops between the probe terminals quantifies the transverse response. 
See the Hamiltonian in Eq.~\eqref{eq:ham-lattice-setup} for details. 
 }\label{fig:aem-lattice}
\end{figure*}

A square lattice with anisotropic nearest-neighbor hoppings is the simplest model that captures an anisotropic bandstructure. However, when the resulting Fermi surface is rotated, forming junctions with other materials becomes difficult. To overcome this, one may consider a two-dimensional square lattice with both nearest-neighbor and next-nearest-neighbor hoppings, which allows the anisotropic bandstructure to be rotated by tuning suitable parameters~\cite{thaku2025}. We first study such a lattice model for an anisotropic bandstructure, as shown in the central region of Fig.~\ref{fig:aem-lattice}. The model can be written as
 \bea 
 H_l &=& -t_x\sum_{n_x,n_y}c^{\dagger}_{n_x+1,n_y}c_{n_x,n_y} -t_y\sum_{n_x,n_y}c^{\dagger}_{n_x,n_y+1}c_{n_x,n_y} \nn \\
 && -t_{+}\sum_{n_x,n_y}c^{\dagger}_{n_x+1,n_y+1}c_{n_x,n_y} \nn \\ &&  -t_{-}\sum_{n_x,n_y}c^{\dagger}_{n_x-1,n_y+1}c_{n_x,n_y}  +{\rm~~ h.c.},  \label{eq:ham-latt}
 \eea 
 where $t_x$ and $t_y$ are nearest-neighbour hopping amplitudes along $x$ and $y$ directions, and $t_+$ and $t_-$ are the next-nearest-neighbour hopping amplitudes along $\hat x+\hat y$ and $\hat x-\hat y$ directions respectively. The operator $c_{n_x,n_y}$ annihilates an electron on site $(n_x,n_y)$. The hopping amplitudes are taken to be real for simplicity. The dispersion relation for this model is given by 
 \bea 
 E &=& -2[t_x\cos{k_xa}+t_y\cos{k_ya}\nn \\ &&~~+t_{+}\cos{(k_x+k_y)a}+t_-\cos{(k_x-k_y)a}] \label{eq:disp-latt}
 \eea
 It can be shown that a Taylor expansion of this dispersion near $k_x = k_y = 0$ reproduces the continuum dispersion (up to an overall energy shift) obtained from the Hamiltonian in Eq.~\eqref{eq:ham-cont}, provided the lattice parameters are chosen such that
 \bea 
 t_x&=&t-\de\cos{2\phi}, ~~t_y=t+\de\cos{2\phi}, \nn \\ 
 && ~~{\rm and}~~t_+=-t_-=-\f{\de\sin{2\phi}}{2} .
 \eea
Physically, the parameter $\delta$ quantifies the anisotropy between the effective hopping amplitudes along two orthogonal directions, while $\phi$ denotes the angle between the principal axes of the anisotropic bandstructure and the transport direction. To estimate realistic values of $\delta$, we relate it to experimentally measured transport anisotropy. From Eq.~\eqref{eq:Gxx}, the longitudinal conductivities at $\phi=0$ and $\phi=\pi/2$ satisfy 
\[
\frac{G_{xx}(\phi=0)}{G_{xx}(\phi=\pi/2)} = \sqrt{\frac{t-\delta}{t+\delta}}.
\]
Using reported anisotropic resistance data for CrSBr, a material known to host a strongly anisotropic bandstructure~\cite{shyaga2026}, we obtain an estimate $\delta \approx 0.43t$, indicating that the parameter regime considered in this work is experimentally realistic.

More broadly, anisotropic hopping parameters of this kind can arise in a variety of systems, including layered van der Waals materials with low crystal symmetry, and strain-engineered  materials~\cite{jiseok,farkous2021} where hopping amplitudes and lattice orientation can be tuned in a controlled manner.

 We now consider a realistic setup in which the transverse response can be experimentally probed. Source and drain terminals are attached to the left and right edges of the lattice hosting the anisotropic bandstructure. In addition, two probe terminals—each modeled as a one-dimensional quantum wire—are connected symmetrically along the transverse (up and down) directions to the same lattice, as illustrated in Fig.~\ref{fig:aem-lattice}. We calculate transverse-voltage using the probe terminals in response to a bias in longitudinal direction using B\"uttiker probe technique~\cite{soori2024deco}. 

\begin{widetext}
The Hamiltonian for the setup shown in Fig.~\ref{fig:aem-lattice} is given by 
\bea 
H &=& H_L+H_M+H_R+H_T+H_B+H_{LM}+H_{MR}+H_{TM}+H_{BM} , \nn \\ 
&& {\rm where}  \nn\\ 
H_L &=& -t\Big[\sum_{n_x=-\infty}^{0}\sum_{n_y=1}^{L_y}c^{\dagger}_{n_x-1,n_y}c_{n_x,n_y}+\sum_{n_x=-\infty}^{0}\sum_{n_y=1}^{L_y-1}c^{\dagger}_{n_x,n_y+1}c_{n_x,n_y}\Big]+{\rm h.c.}-\mu_L\sum_{n_x=-\infty}^{0}\sum_{n_y=1}^{L_y} c^{\dagger}_{n_x,n_y}c_{n_x,n_y} , \nn \eea
\bea H_M &=& -t_x\sum_{n_x,n_y}c^{\dagger}_{n_x+1,n_y}c_{n_x,n_y} -t_y\sum_{n_x,n_y}c^{\dagger}_{n_x,n_y+1}c_{n_x,n_y}  -t_{+}\sum_{n_x,n_y}c^{\dagger}_{n_x+1,n_y+1}c_{n_x,n_y}  -t_{-}\sum_{n_x,n_y}c^{\dagger}_{n_x-1,n_y+1}c_{n_x,n_y}  \nn\\ 
&& +{\rm~~ h.c.}-\mu_M\sum_{n_x=1}^{L_x}\sum_{n_y=1}^{L_y} c^{\dagger}_{n_x,n_y}c_{n_x,n_y}, \nn\\
H_R&=& -t\Big[\sum_{n_x=L_x+1}^{\infty}\sum_{n_y=1}^{L_y}c^{\dagger}_{n_x+1,n_y}c_{n_x,n_y}+\sum_{n_x=L_x+1}^{\infty}\sum_{n_y=1}^{L_y-1}c^{\dagger}_{n_x,n_y+1}c_{n_x,n_y}\Big]+{\rm h.c.} -\mu_R\sum_{n_x=L_x+1}^{\infty}\sum_{n_y=1}^{L_y} c^{\dagger}_{n_x,n_y}c_{n_x,n_y}, \nn
\eea
\bea
H_T&=&-t\sum_{n=1}^{\infty}(d^{\dagger}_{n+1}d_n+{\rm h.c.} )-\mu_P\sum_{n=1}^{\infty}d^{\dagger}_{n}d_n, ~~~~
H_B~=~-t\sum_{n=-\infty}^{-1}(d^{\dagger}_{n-1}d_n+{\rm h.c.} )-\mu_P\sum_{n=-\infty}^{-1}d^{\dagger}_{n}d_n, \nn\\
H_{LM} &=& -t'\sum_{n_y=1}^{L_y}( c^{\dagger}_{0,n_y}c_{1,n_y} +{\rm h.c.}), ~~~~~~~~~~~~~~~~~
H_{MR} ~=~ -t'\sum_{n_y=1}^{L_y} (c^{\dagger}_{L_x,n_y}c_{L_x+1,n_y} + {\rm h.c.}), \nn \\
H_{TM}&=&-t_P(d^{\dagger}_1c_{n_{x0},L_y}+{\rm h.c.}),~~~~~~~~~~~~~~~~~~~~~~
H_{BM}~=~-t_P(d^{\dagger}_{-1}c_{n_{x0},1}+{\rm h.c.}) ,~~ \label{eq:ham-lattice-setup}
\eea
\end{widetext}
where $H_L$ ($H_R$) is the Hamiltonian for the left (right) lead, $H_M$ is the Hamiltonian for the central lattice that hosts anisotropic bandstructure, $H_T$ ($H_B$) is the Hamiltonian for the top (bottom) probe, $H_{LM}$ ($H_{MR}$) connects the left (right) lead to the central lattice, and $H_{TM}$ ($H_{BM}$) connects the top (bottom) probe to the central lattice. 
Here, $\mu_L$ and $\mu_R$ denote the chemical potentials of the source and drain terminals, respectively, and are taken to be equal, $\mu_L = \mu_R = \mu_n$. The central square lattice with anisotropic bandstructure has chemical potential $\mu_M$. The operator $d_n$ annihilates an electron on site $n$ of probe terminal. The hopping amplitude connecting the anisotropic central lattice to the left and right terminals is denoted by $t'$, while the hopping amplitude linking the central lattice to the probe terminals is $t_P$.

\begin{figure}
\includegraphics[width=7cm]{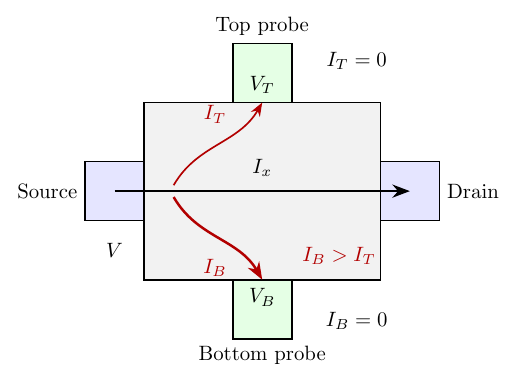}
\caption{Schematic illustrating the B\"uttiker probe technique. The sample is connected to source and drain terminals along the longitudinal direction, and to top and bottom probe terminals in the transverse direction. A bias applied between source and drain drives a longitudinal current, which is deflected asymmetrically toward the transverse probes due to anisotropy. The probe voltages $V_T$ and $V_B$ adjust self-consistently such that the net current through each probe vanishes. The resulting voltage difference $V_H = V_T - V_B$ defines the transverse voltage.}\label{fig:multiterminal}
\end{figure}
We now briefly outline the method of calculation and provide an intuitive picture of the transverse response within the B\"uttiker probe framework. In a multiterminal setup [see Fig.~\ref{fig:multiterminal}], the probe terminals are defined such that the net current flowing into them is zero.  The probe voltages adjust self-consistently to ensure zero net current in the transverse terminals, thereby preventing charge accumulation. When the system exhibits an imbalance in the flow of carriers toward the two transverse directions, the probe voltages $V_T$ and $V_B$ shift relative to each other to compensate for this imbalance, resulting in a finite transverse voltage.

We compute this quantitatively using scattering theory. The scattering eigenfunction corresponding to an electron incident in the $m_0$-th channel of the source at energy $E$ is first obtained. This eigenfunction is used to evaluate the currents flowing into the probe terminals. The total current through each probe terminal due to a longitudinal bias $V$ is obtained by summing over all incident channels within the bias window. 

Next, scattering eigenfunctions corresponding to electrons incident from each of the probe terminals are computed, and the associated currents are evaluated for a range of probe voltages $(V_T, V_B)$. The total currents in the probe terminals therefore include contributions from both the source--drain bias $V$ and the applied probe voltages. The values of $(V_T, V_B)$ are then determined by imposing the condition that the net current in each probe terminal vanishes. The resulting voltage difference $V_T - V_B$ defines the transverse voltage $V_H$. All calculations are performed numerically.

\begin{figure}
\includegraphics[width=4.2cm]{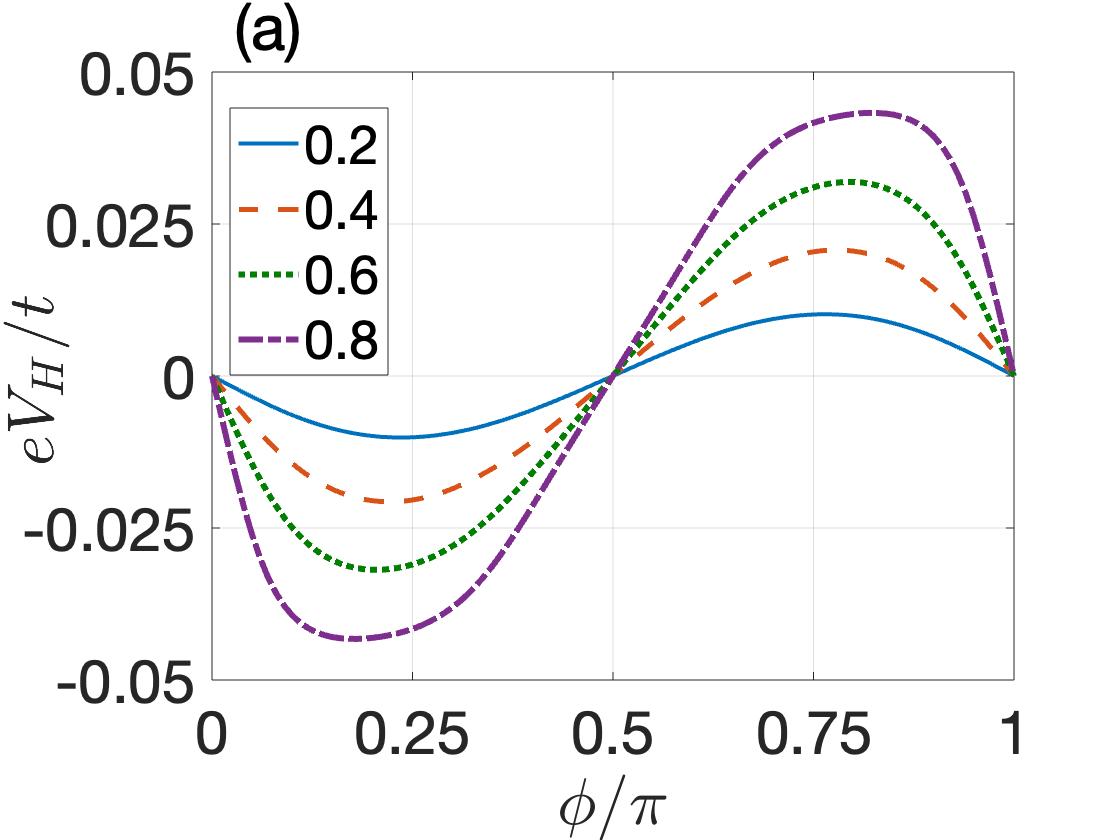}
\includegraphics[width=4.2cm]{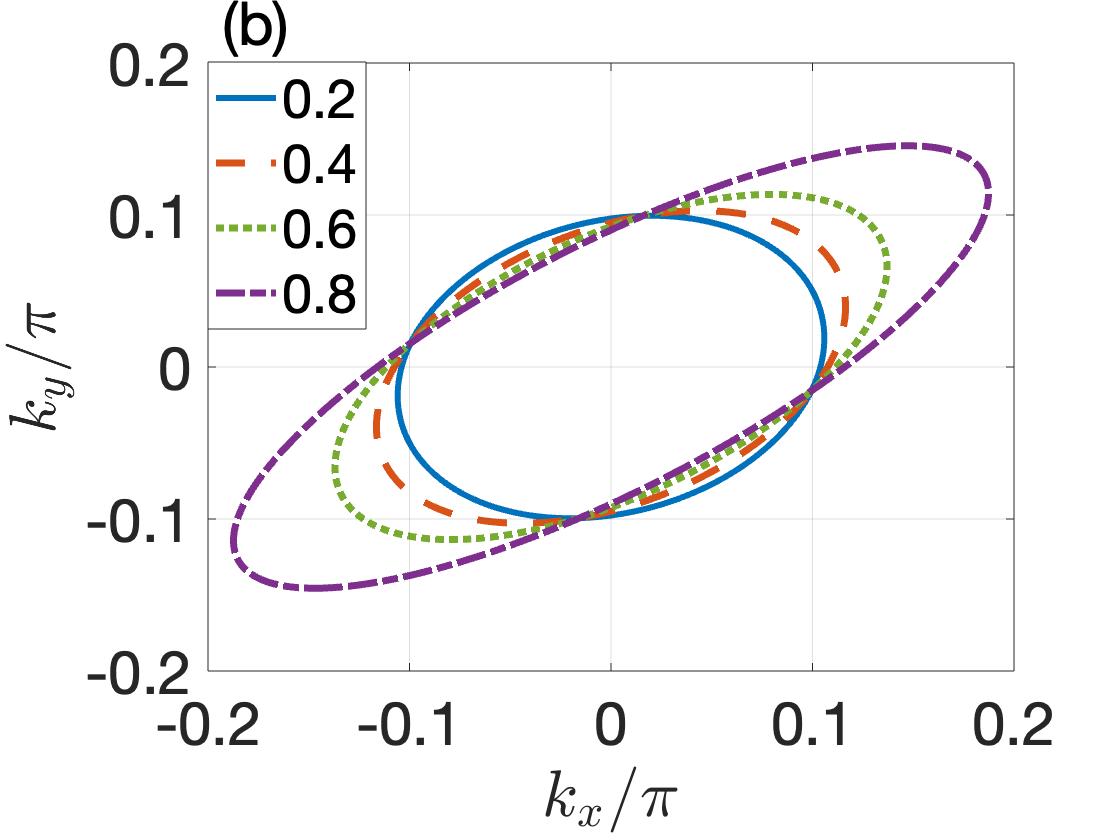}
\includegraphics[width=7cm]{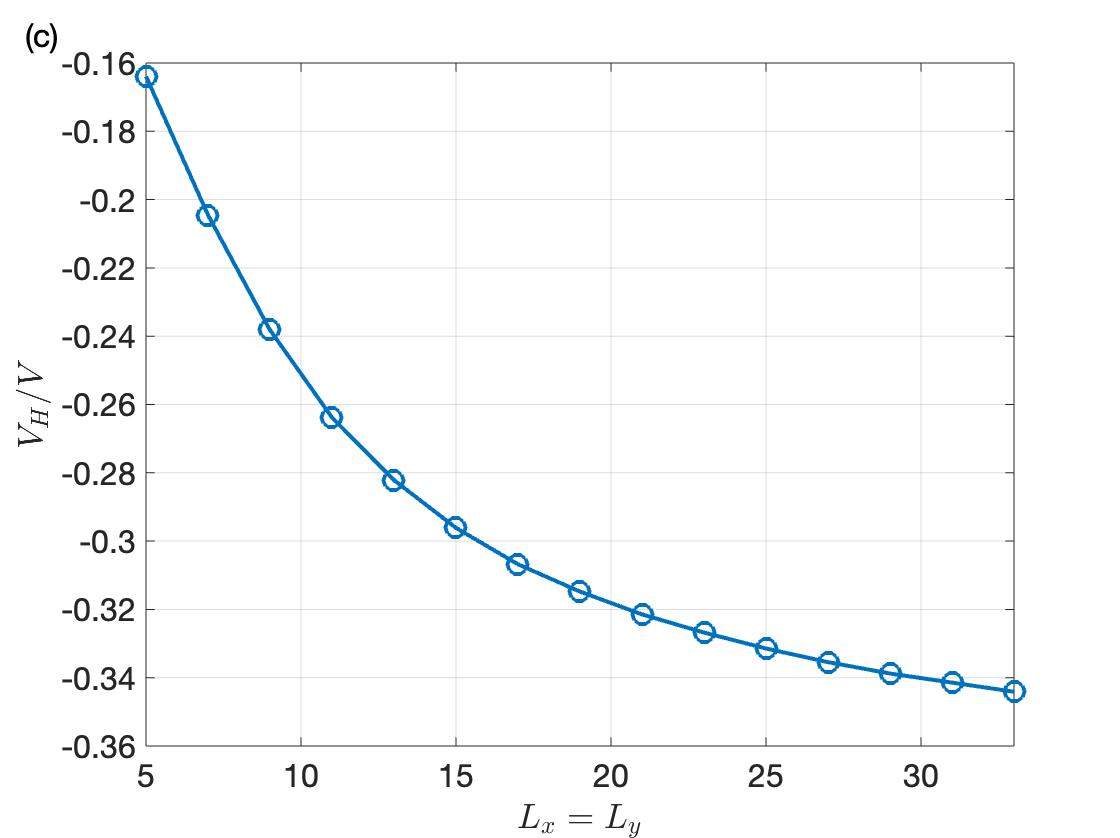}
\caption{ (a) Transverse voltage as a function of $\phi$, the angle of rotation of the system. Parameters: 
$L_x=L_y=11$, $n_{x0}=6$, $\mu_n = -2t$, $\mu_M = -4t$, $t' = t$, $t_P = 0.05t$, and $eV=0.05t$. 
Different curves correspond to different values of $\delta$. 
(b) Fermi contour of the central lattice for various choices of $\delta$ at energy $0.1t$. 
The legends indicate the values of $\delta/t$. (c) Hall voltage versus system size $L_x=L_y$ for fixed values of $\phi=\pi/4$, $eV/t=10^{-5}$. The Hall voltage is calculated at the center, i.e., $n_{x0}=(L_x+1)/2$. }~\label{fig:results-lattice}
\end{figure}

In Fig.~\ref{fig:results-lattice}(a), we plot the transverse voltage $V_H$ as a function of the rotation angle $\phi$ for 
$L_x = L_y = 11$, $n_{x0} = 6$, $\mu_n = -2t$, $\mu_M = -4t$, $t' = t$, $t_P = 0.05t$, and $eV = 0.1t$. 
As the anisotropy parameter $\delta$ increases, the magnitude of the transverse voltage grows, 
since the symmetry $k_y \to -k_y$ is increasingly broken at larger $\delta$. 
Furthermore, at $\phi = 0, \pi/2, \pi$, the transverse voltage vanishes because the Fermi surface regains symmetry under 
$k_y \to -k_y$. 
Figure~\ref{fig:results-lattice}(b) shows the Fermi contour of the central lattice at $\phi = 0.3\pi$ for different values of $\delta/t$.

To examine finite-size effects, in Fig.~\ref{fig:results-lattice}(c) we show the scaling of the transverse voltage $V_H$ with system size. We fix the applied bias and compute $V_H$ as a function of $L_x = L_y$, evaluating it at the central site $n_{x0} = (L_x+1)/2$. We find that the magnitude of $V_H$ does not show systematic suppression with increasing system size, indicating that the effect is not a finite-size effect.

Importantly, the angles $\phi$ at which the transverse response vanishes remain unchanged as the system size is increased. This demonstrates that the symmetry-protected zeros of $V_H$ are robust and not affected by finite-size effects.

The dependence of the transverse response on the Fermi energy can be understood analytically within the continuum model. As shown in Eq.~\eqref{eq:Gyxc}, the transverse conductivity scales as $\sqrt{E}$, reflecting the energy dependence of the density of states and the underlying anisotropic dispersion. This indicates that the transverse response varies smoothly with carrier density and can, in principle, be tuned continuously by electrostatic gating in experiments. A similar qualitative dependence shows up in the lattice model, since it reproduces the continuum dispersion near the band bottom.
Although relatively large biases are used in the numerical calculations for clarity, the transverse voltage scales linearly with the applied bias.

It is instructive to contrast this behavior with that of conventional Hall effects. In the quantum Hall effect, the Hall conductivity exhibits quantized plateaus as a function of Fermi energy due to Landau level formation in the presence of a magnetic field.  In contrast, the transverse response in the present work arises purely from Fermi-surface anisotropy and varies continuously with energy, without requiring magnetic fields.

\section{Discussion}

The emergence of a transverse response in our system can be understood from symmetry considerations. In an isotropic system or when the principal axes of an anisotropic Fermi surface are aligned with the transport direction, the dispersion is symmetric under $k_y \to -k_y$. As a result, states with opposite transverse momenta contribute equally and oppositely to the transverse current, leading to its cancellation. However, when the anisotropic Fermi surface is rotated with respect to the transport axes, this mirror symmetry is broken, and the cancellation no longer holds, resulting in a finite transverse response. 

The transverse conductivity therefore vanishes at special orientations $\phi = n\pi/2$ (with integer $n$), where the mirror symmetry $k_y \to -k_y$ is restored. We emphasize that this effect does not involve breaking of time-reversal symmetry. 

Onsager reciprocity relations apply to the full multiterminal conductance matrix $G_{pq}$, which satisfies $G_{pq}(B) = G_{qp}(-B)$ (where $B$ is magnetic field) and is symmetric at zero magnetic field. Here, the subscripts $p, q$ refer to the indices that represent terminals and $G_{pq}$ refers to the differential ratio of current in terminal $p$ to the voltage in terminal $q$. On the other hand, $G_{yx}$ calculated in our work refers to the differential ratio of current in the $y$-direction to voltage applied in the $x$-direction. The transverse response reported in our work using lattice model is obtained after imposing probe conditions (zero current in the transverse terminals) and therefore corresponds to an effective response coefficient rather than a direct element of the conductance matrix. As a result, it is not constrained to satisfy a simple symmetry under index exchange.

The transverse response discussed here originates from the bulk anisotropy of the dispersion and is therefore expected to be robust against variations in contact properties, although the precise magnitude of the measured voltage may depend on contact details.

Examples of materials exhibiting strongly anisotropic bandstructures include quasi-two-dimensional organic conductors such as $\beta''$-(BEDT-TTF) salts~\cite{salameh2007,alemany2015}, van der Waals materials such as CrSBr~\cite{shyaga2026}, and low-symmetry transition metal dichalcogenides such as ReSe$_2$~\cite{lin2015,liu2020}. In particular, CrSBr has recently attracted significant attention due to its pronounced in-plane anisotropy arising from its low-symmetry crystal structure~\cite{shyaga2026}, making it a promising platform for realizing the mechanism discussed in this work. These systems provide experimentally accessible platforms where the orientation of anisotropic electronic structure relative to transport direction can be controlled, for example via device geometry or strain.

The results presented here are obtained in the ballistic regime. The underlying mechanism, however, originates from the anisotropy of the Fermi surface and is therefore expected to persist in the presence of weak to moderate disorder.  Disorder primarily affects the magnitude of the response, and only in the limit of strong disorder the transverse response can be  suppressed. 

In the low-filling regime, the lattice model reproduces the continuum dispersion near the band bottom, and therefore the two approaches yield qualitatively consistent results. A direct quantitative comparison between the two approaches is, however, not straightforward, since the continuum calculation yields a transverse conductivity in a translationally invariant system, whereas the lattice calculation determines a transverse voltage in a finite multiterminal geometry using the B\"uttiker probe framework. 

\section{Conclusion} 
We have predicted a transverse response in electron transport through a metal with an anisotropic 
bandstructure, even in the absence of a magnetic field or Berry curvature. Using a two-dimensional continuum 
model that is translationally invariant in both directions, we demonstrated that an anisotropic and rotated 
Fermi contour naturally leads to a finite transverse conductivity.

To connect this continuum picture to a realistic mesoscopic setup, we constructed a corresponding lattice 
model with direction-dependent nearest- and next-nearest-neighbor hoppings. This lattice model faithfully 
reproduces the desired continuum dispersion and, importantly, allows controlled rotation of the anisotropic 
bandstructure by tuning a small set of microscopic parameters. We then attached source and drain terminals 
along the longitudinal direction and voltage-probe terminals along the transverse direction, forming a 
multiterminal transport geometry suitable for numerical study. Using the Büttiker-probe method, we computed 
the transverse voltage generated across the transverse probes and found  that it qualitatively agrees with the 
transverse conductivity obtained from the continuum model.

Our results show that the magnitude of the transverse response is directly governed by the degree of anisotropy 
in the Fermi contour and its lack of symmetry under $k_y \to -k_y$. The response vanishes at specific rotation 
angles where this symmetry is restored. These features provide clear experimental signatures that can be tested 
in mesoscopic devices where   strain can be possibly used to tune anisotropy. Importantly, unlike 
the quantum Hall effect~\cite{Klitzing1980}, the transverse response predicted here is \emph{not quantized}; rather, it is a continuous 
function of the anisotropy and the rotation angle of the Fermi surface.

Altermagnets provide a natural platform featuring intrinsically anisotropic bandstructures~\cite{smejkal2022,Song2025}. 
However, the dispersions for the two spin species in altermagnets are rotated by $90^{\circ}$ relative to each other 
in the $k_x\!-\!k_y$ plane. Consequently, to access the effects presented in this work, one must use ferromagnetic 
source and drain electrodes so that transport is dominated by electrons of a single spin species. Thus, our predictions 
provide an experimentally relevant transport  probe of altermagnetic anisotropy, potentially complementing spectroscopic studies.

Beyond altermagnets, our work highlights a broader principle: anisotropy combined with Fermi-surface rotation 
can mimic aspects of Hall physics without requiring magnetic fields or topological Berry-curvature effects. 
This opens a route to engineer transverse signals in a wide class of low-symmetry materials, including strained 
metals, anisotropic two-dimensional materials, and artificial lattices.

In summary, we have shown that anisotropic bandstructures with tunable orientation lead to a robust and 
measurable  transverse response. Our study provides a unified continuum and lattice framework to 
analyze this phenomenon, identifies its symmetry origins, and proposes realistic device geometries for its 
experimental detection. We hope our findings stimulate further exploration of anisotropy-driven transport 
effects in emergent materials.

\begin{acknowledgements}
The author thanks  Gajanan V Honnavar, Manisha Thakurathi, Dhavala Suri, R Ganesh,  Sreejith G J and Subroto Mukerjee  for useful discussions.  
The author thanks Science and Engineering Research Board (now Anusandhan National Research Foundation) - Core Research grant (CRG/2022/004311) and University of Hyderabad for financial support. 
\end{acknowledgements}

\bibliography{refaem}

\end{document}